# The study of synchronous (by local time) changes of the statistical properties of thermal noise and alpha-activity fluctuations of a $^{239}$Pu sample.

A.V. Kaminsky, S.E. Shnoll

Lomonosov's Moscow State University, Moscow, Russia
National High Technology Centre, Tbilisi, Georgia

**Introduction**

Let us consider a random process $X(t)$. It can be the fluctuation noise of current in a resistor, a series of measurements of the rate of a chemical reaction, the rate of photon counting with a photomultiplier, the radioactive decay or any other process generating the random signal $X(t)$. It is easy to obtain the one-dimensional probability density $\rho(X)$ of the quantity $X$. This can be achieved by measuring the relative time of the signal to be between $X$ and $(X + \Delta X)$ within a large enough interval $T$. Then the distribution function $\rho(X)$ will verge towards the normal distribution. Being an integral characteristic of a stationary, in wide sense, process, this function will certainly not depend on the current time. Let us imagine now that the measurement longs a certain fixed time $t$, $t + \tau$ ($\tau \ll T$), which is not large enough to satisfy either criterion of statistical validity. Then we get a statistically inconsistent exemplar of the distribution $G(X, n) \neq \rho(X)$, changing from realization to realization. And the object of our study will be an experimentally obtained series of functions $G(X, n)$ represented in a form of histograms for non-overlapping intervals $t_0 + n\tau$, $t_0 + (n + 1)\tau$.

A visual evaluation of the similarity of histogram patterns in the time series $G(X, n)$ reveals the following interesting laws:

1) the effect of "near zone"- a much higher probability to be similar for the histograms constructed from the nearest, neighboring non-overlapping segments of a time series of the results of measurements;
2) the local time effect - high probability of similar histograms to appear synchronously by local time, this being observed upon measuring process of various nature at different geographical points;
3) the existence of two daily periods: "sidereal" (1436 min) and "solar" (1440 min).

All these effects are unified by a common term "macroscopic fluctuations" (MF) and mean regular-in-time changes of the fine structure of the spectrum of amplitude fluctuations (i.e. changes of the pattern of the corresponding histograms) in the process of various nature [1- 8].
The major results in the study of macroscopic fluctuations have been obtained by measuring the alpha-activity of $^{239}$Pu samples. In spite of having a number of advantages (relating to the stability towards ordinary physical perturbations), this object has a certain drawback: it does not allow us to conduct measurements with high time resolution. To do this, one has to use samples with a very high activity, which is not compatible with the "dead time" of the corresponding detectors.

The measurements on shorter time scales would be especially interesting in regard to the effect of near zone. The high probability of histograms to be similar in the non-overlapping nearest segments of a time series means that there is an external cause, a factor that determines the histogram pattern. Hence, it would be natural to try and find the "lifetime" of the "Idea of a certain Form" by constructing histograms for more and more short intervals. It could be expected that the size of the near zone (in the number of successive intervals) would grow at such a decrease of the interval duration. However, the lower time limit for radioactivity measurements turned out to be 0.1-1 s.

One of the tasks of our research was the study of the effect of near zone with the highest possible time resolution. This is necessary for elucidation of the lower boundary of the characteristic time scale of MF-factor.

Another task was to reproduce the main results obtained earlier (see items 1-3) for process of different nature. The goal was either to confirm or disprove the supposition that MF-factor is not selective towards different physical processes and, hence, will act at a quite deep level of universe organization, probably affecting the metric properties of the space-time.

**Methods, devices and analysis of results**

As the first source of random numbers, we used a well-elaborated method, which is based on the measurement of alpha-activity of $^{239}$Pu samples. The generator was set up in Pushchino (Moscow Region, Russia). Another generator was placed in Tbilisi (Georgia) and represented a two-channel device, which amplified ($k \sim 10^6$) and digitized equilibrium thermal noise (Johnson noise) from two metal-film resistors, operating at room temperature. The fluctuations of electrical charges in resistance is a universal phenomenon, which determines the lower boundary of noise in any source of signal [9]. Johnson noise voltage created by a resistor, operating at a temperature $T$, is determined by the known Nyquist formula, which follows from basic thermodynamical considerations:

(1) $$U_f = \sqrt{4kTR\Delta f}$$

where $k$ is the Boltzmann constant, $T$ is temperature, $R$ is resistance, $\Delta f$ is a frequency band. The operation speed of ADC was 8 kHz. However, for data to be compatible with the time series obtained from the radioactivity measurements, only the first value from 8000 samples was rounded to the nearest integer and placed into the computer archive. In addition, the value of standard deviation was calculated for the 8000-point array once per second and stored in the archive. The signals were recorded on two independent channels. The experiment started on June 4, 2005 at 00:00 by Moscow winter time.
In the experiment, the following time series were obtained.

In Pushchino:
$X1_{Push}(t)$, the rate of alpha-decay. Every member of the series represents the number of impulses on the detector of alpha-particles counted for the previous second. This series was obtained on a setup with a collimator, directing a narrow beam of alpha-particles along the emitter-detector axis oriented west;
$X2_{Push}(t)$, an analogous series obtained on a setup with a rotating collimator [7], always oriented at the Sun.

In Tbilisi:
$X1_{Tbil}(t)$, the equilibrium noise of the 1$^{st}$ resistor. Every member of the series represents the voltage on the terminals of resistor at the beginning of each second.
$S1_{Tbil}(t)$, the corresponding series of standard deviations for $X1_{Tbil}(t)$.
$X2_{Tbil}(t)$, the equilibrium noise of the 2$^{nd}$ resistor. Every member of the series represents the voltage on the terminals of resistor at the beginning of each second.
$S2_{Tbil}(t)$, the corresponding series of standard deviations for $X2_{Tbil}(t)$.

The time series obtained were processed according to the procedure described in details in [4-6].

**Results**

Obtained from measuring both the voltage noise in a resistor or radioactivity, the time series represent "white noise" without any noticeable trends (Fig. 1). In the case of radioactivity, it is a typical process that obeys Poisson statistics. The analysis performed shows no cross-correlation between the time series obtained, with a good δ-correlation of each single signal. This allows us to conclude that the successive measurements are statistically independent of each other.

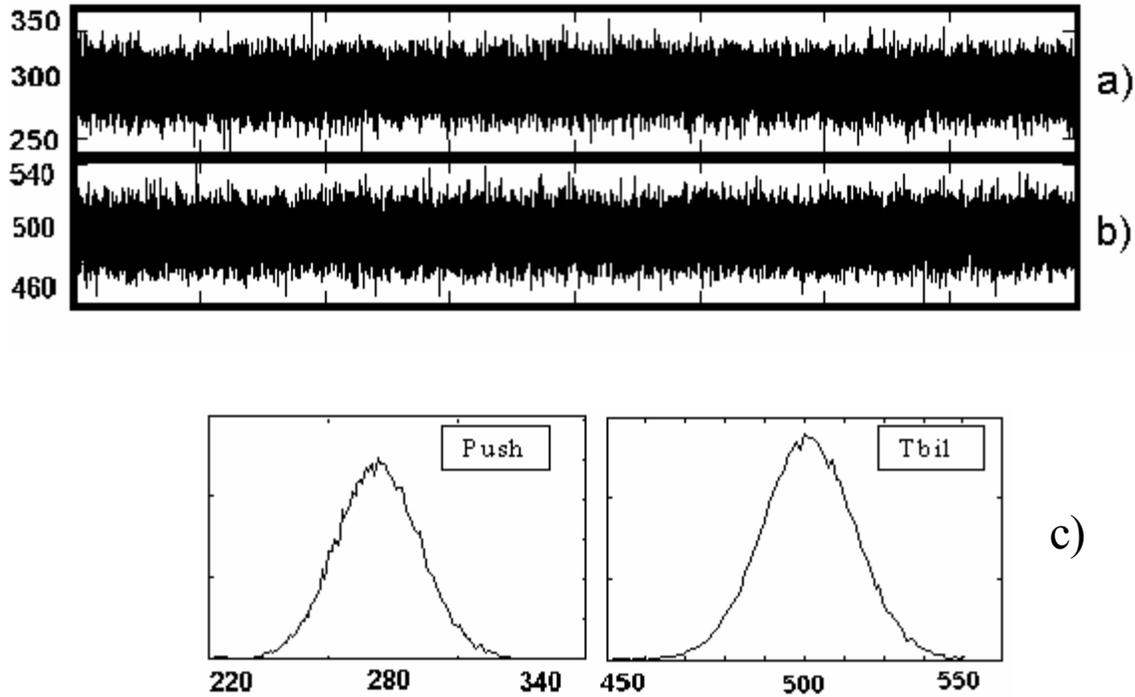

**Fig. 1a. The time series $X1_{Tbil}(t)$, 1$^{st}$ Tbilisi array. A 10-h record fragment. Plotted along X-axis is time. Interval between measurements is 1 s. Plotted along Y-axis is noise voltage on the resistor terminals expressed in relative units.**

**Fig. 1b. The time series $X_{Push}(t)$, the results of measurements of $^{239}$Pu alpha-activity in Pushchino. Plotted along X-axis is time. Interval between measurements is 1 s. Plotted along Y-axis is activity in cps.**

**Fig. 1c. Distribution functions for the corresponding signals $\rho(X_{Push})$ и $\rho(X1_{Tbil})$ constructed on the whole ensemble of realizations (172800 samples).**

Fig. 2 shows first 12 smoothed histograms for the time series obtained in Tbilisi and Pushchino during a synchronous experiment. The processing of the experimental data comes to the comparison of the histogram arrays. The histograms are obtained by a sevenfold moving averaging over 5 points. To get significant results and to construct distributions of the number of similar pairs over the duration of the interval between them, one should do about 10000 comparisons (for every variant of the task).

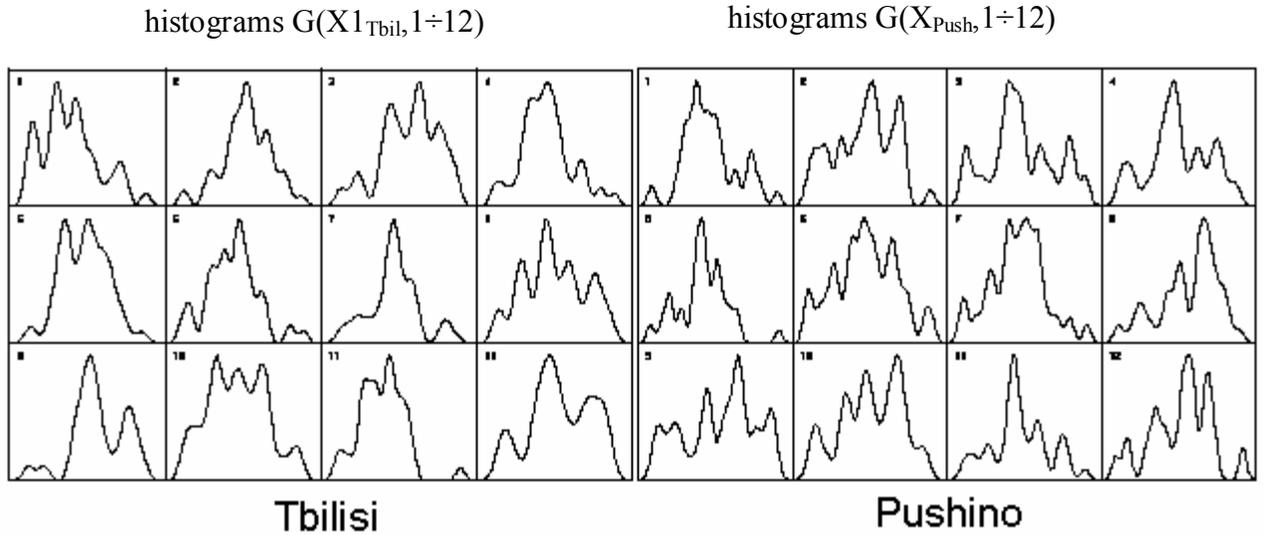

**Fig. 2. Fragments of the computer journal. The left diagram shows 1-min histograms constructed from 60 one-second measurements of the resistor electric noise in Tbilisi. The right diagram shows 1-min histograms constructed from 60 one-second measurements of the $^{239}$Pu alpha-activity in Pushchino. The order numbers of histograms (as they are in the histogram successive series) are given.**

Fig. 3 shows the results of comparison of histograms constructed from the measurements of the resistor electric noise in Tbilisi. As can be seen from the figure, there is a quite clear effect of the near zone.

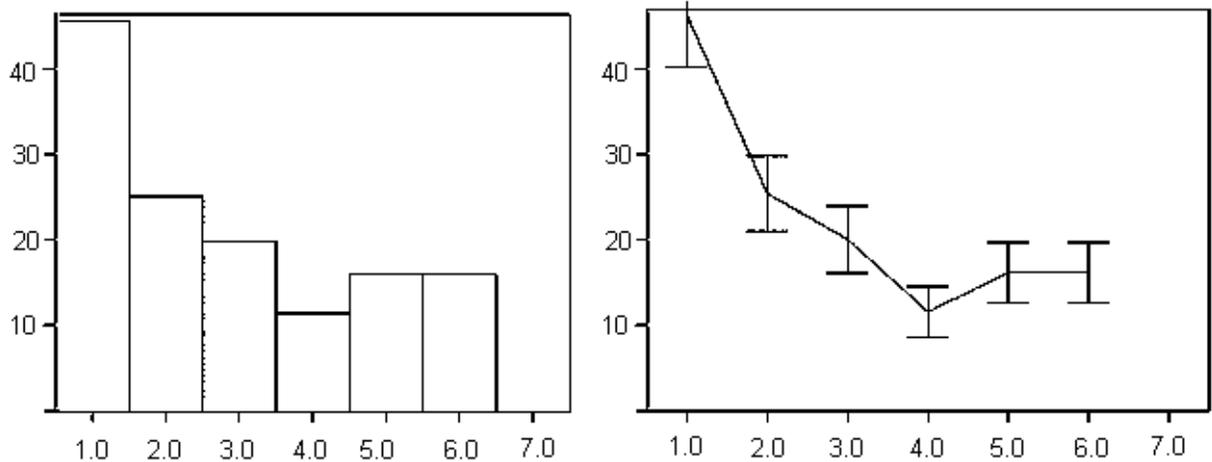

**Fig. 3. The effect of near zone: the number of similar pairs among 1-min histograms depending on the duration of the interval between them. The histograms are constructed from 60 one-second measurements of noise taken from Tbilisi array no. 1. Plotted along abscissa axis are time intervals ($t_1 - t_2$) in min. Ordinate is the number of matches between histograms from $G(X1_{Tbil}, t_1)$ и $G(X1_{Tbil}, t_2)$, which corresponds to an interval ($t_1 - t_2$).**

Fig. 4 shows the result of comparison of histograms constructed from the noise signals of two independent resistors. The resistors were half a meter apart, and all circuits were safely shielded. The absence of electrical or any other connection between the channels was verified by calculation of the cross-correlation function. As can be seen from Fig. 4, there is a kind of informational relation between the channels revealed by the method of histogram comparison. The relation, which is expressed in the similarity of the patterns of current inconsistent distributions, exists despite the absence of a physical connection between the channels (the mutual energy of the channels is equal to 0).

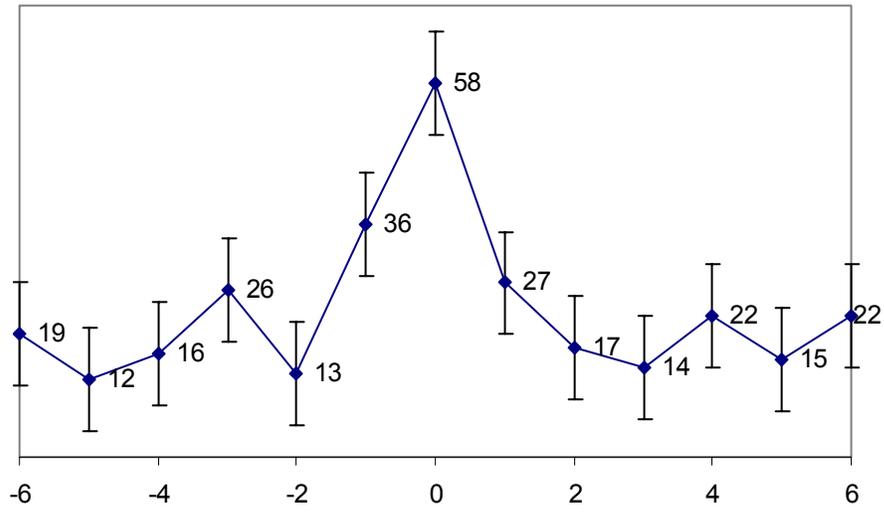

**Fig. 4. As follows from the measurement of noise in two independent generators in Tbilisi, the probability of similar histograms to be realized is maximal at the same time moment. Plotted along abscissa axis is the time interval ($t_1 - t_2$, in min) between the histograms under comparison, $G(X1_{Tbil}, t_1)$ and $G(X2_{Tbil}, t_2)$. The zero interval corresponds to the absence of delay. The numerals on the chart indicate the number of similar histograms for the given interval ($t_1 - t_2$).**

Fig. 5 shows an analogous curve obtained upon comparison of the histograms constructed from a series of the resistor voltage fluctuations and the corresponding series of standard deviations. There is a clear correlation between the histograms realized synchronously.

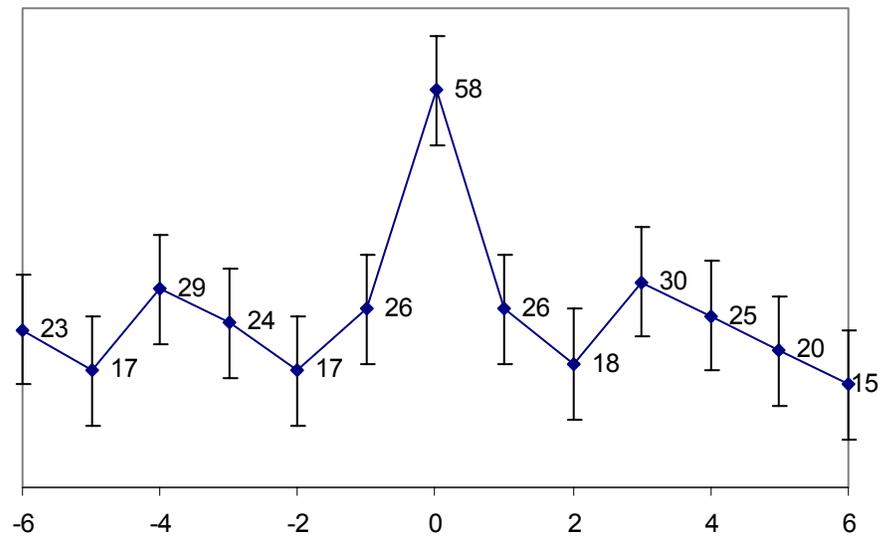

**Fig. 5. The correlation in the occurrence of similar histograms constructed from the series of the resistor voltage fluctuations, $G(X1_{Tbil}, t_1)$, and the corresponding series of standard deviations, $G(S1_{Tbil}, t_2)$.**

The next figure demonstrates a local-time synchronization of the appearance of similar histograms in Tbilisi and Pushchino. By absolute time, similar histograms appear in Pushchino 28 min later, and this value is not accidental. It exactly corresponds to the local time difference, which takes into account the geographical coordinates of Pushchino and Tbilisi.

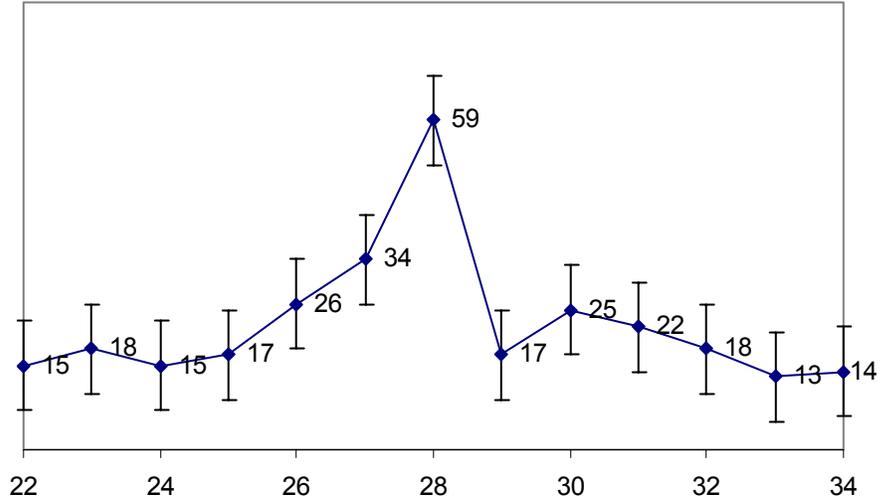

**Fig. 6. Similar histograms constructed from measurements of the resistor electric noise in Tbilisi and the $^{239}$Pu alpha-activity in Pushchino are highly probable to appear synchronously by local time. The calculated difference in local time is 28.8 min. Plotted along abscissa axis are time intervals ($t_1 - t_2$) between the histograms under comparison, $G(X1_{Tbil}, t_1)$ и $G(X2_{Push}, t_1)$; along ordinate axis is the number of matches found.**

The search for similar histogram patterns among pairs, where one histogram is delayed from another by about a day, reveals two maxima (with the time differences between histograms of 1436 and 1440 min). The maximum of 1436 min corresponds to the period of the Earth rotation relative to the immobile stars (the sidereal day), while the maximum of 1440 min represents the ordinary solar day. These maxima were also invariably found in all our earlier experiments with the radioactive generators of random numbers.

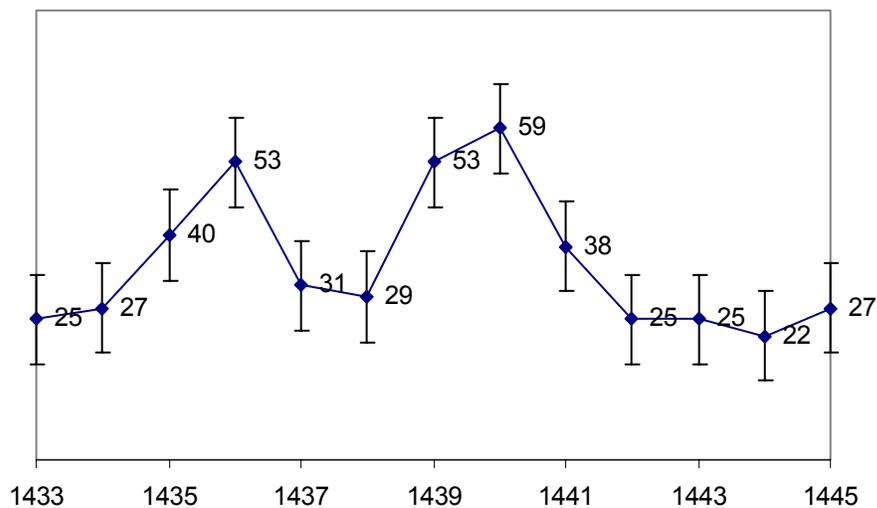

**Fig. 7. The "solar" and "sidereal" daily periods in the change of the probability of similar histograms to appear. The histograms are constructed from 60 one-second measurements of the resistor equilibrium heat noise in Tbilisi. Plotted along abscissa axis are intervals $t_1 - t_2$, along ordinate axis are the number of matches for a given interval found upon comparing $G(X1_{Tbil}, t_1)$ and $G(X1_{Tbil}, t_2)$.**

**Discussion and conclusions**

In the main, the results of the present work reproduce the results obtained earlier upon studying macrofluctuations. All these effects (the effect of near zone, the existence of solar and sidereal cycles in the appearance of similar histogram patterns etc.) were already demonstrated more than once in numerous experiments. What is new is the demonstration of these effects for a new object, the equilibrium electrical noise in the resistor, and also the proof of MF-factor being insensitive to the nature of the physical processes, generating the stream of stochastic events. The latter is evidenced by the experiment described in the paper: the patterns of histograms obtained in Pushchino upon the analysis of the alpha-decay rate correlate with those obtained in Tbilisi upon the analysis of the noise signal of electrical fluctuations in the resistor. Thus, processes, which cannot have any relation from the traditional point of view, turn out to be related when analysed macroscopic fluctuations. By now, there is no a satisfactory model, which could explain this phenomenon. It is clear, however, that non-ordinariness of the phenomenon would require a non-ordinary approach to its explanation. At present, we can speak about two qualitative models.

The first model implies the existence of an external cosmo-physical factor [2, 4, 5, 8], affecting physical processes and resulting in a wide spectrum of phenomena which unified by the term "macroscopic fluctuations". Therefore, there may be a stable field structure, a kind of a 3D- lattice, which is formed in space by a wave field upon the reflection from the walls of a complex-shaped resonator. The movement of the Earth together with our apparatus through such a lattice could well explain the discovered cosmo-physical cycles and a number of other features of the MF-phenomenon. However, unclear is the nature of this field, whose very specific effect on the statistics of stochastic processes defies simple explanations.

Another approach is more radical and implies a fundamental change in our views on probability and determinacy. This approach is based on novel ideas in the theory of quantum mechanics [10,11] and suggests a latent relation of phenomena (MF-factor?) on the basic level of reality, which is inaccessible to the direct observation. This means that any phenomena, which are stochastic on the level of physical reality, can be non-stochastic on a deeper level of matter organization. Determinism on the basic level of matter organization leads us to the conception of the world as a finite automaton with a vast number of states. It is important to realize the fact that we together with our apparatus are a part of this world. The dynamics of such a "world automaton" in the space of its states can approach, under some conditions, a discrete analog of deterministic chaos. An example of the simplest system, generating deterministic chaos, is the quadratic map that describes the dynamics of populations in a bioniche [12]:

(2) $$x_{n+1} = 4rx_n(1-x_n)$$

It is known that the behavior of such a system will be determined by the parameter $r$. At $r = r_c \to 0.892...$, the behavior of the system becomes chaotic after a number of bifurcations. How would the world dynamics appear to an observer, who is a part of this model world? For the case under consideration, which is quite simple, the answer will not be difficult. We should just take into account that the observer will take a part of the computational resources (cells) of the world automaton. Obviously, he will not have access to all states of the world. The Nature would appear to him much more rough and coarse than it is in reality. In other words, the observer will see the world renormalized. The fractal chaos will gain features of an order. So if in our model, the world is near the critical state $r_c$, then after the scale transformation, the value of the parameter $r$ will diminish, and we will fall into a bifurcational region, where a certain structural order rules in a multi-stage cycle. We suppose that cyclic processes, from subatomic oscillations to the dynamics of galactic clusters, are the consequence of our renormalized subjective point of view. The observed fractal similarity of the world also confirms this hypothesis. It is important that in such a model, all processes are interconnected by a unified algorithm, similar to the recurrent formula (2). And this would give a clue to the understanding of MF-processes. The stochastic – from the viewpoint of traditional statistical criteria – character of histogram patterns can actually result from the dynamics of hidden parameters at a basic level of matter organization [13,14]. In conclusion, let us express a hope that the experimental investigation of the fine structure of fluctuations, including the experiments described in this paper, will enable us to understand the nature of the latent MF-factor in the nearest future.

**References**
1. Shnoll S.E. Macroscopic fluctuations with a discrete distribution of amplitudes in the processes of various nature. In "Advances in Science and Engineering. General Problems of Physico-Chemical Biology", v. 5, p. 130-201. (1985)
2. S.E.Shnol', N.V.Udaltzova and N.B.Bodrova "Macroscopic fluctuations with discrete structure distributions as a result of universal causes including cosmophysical factors In: Proc.First Intern. Congress on Geo-cosmic Relations. Wageningen, Netherlands pp.181-188 (1989)
3. S E Shnoll, V A Kolombet, E V Pozharskii, T A Zenchenko, I M Zvereva, A A Konradov "Realization of discrete states during fluctuations in macroscopic processes" , Physics-Uspekhi **41**,(10), 1025-1035 (1998)
4. S.E.Shnoll , E.V.Pozharski , T.A.Zenchenko , V.A.Kolombet , I.M.Zvereva , A.A.Konradov "Fine Structure of Distributions in Measurements of Different Processes as Affected by Geophysical and Cosmophysical Factors» Phys. Chem. Earth (A), Vol.24, No.8, pp. 711-714, (1999)
5. S E Shnoll, T A Zenchenko, K.I.Zenchenko, E V Pozharskii, V A Kolombet, A A Konradov "Regular variation of the fine structure of statistical distributions as a consequence of cosmophysical agents" Physics – Uspekhi **43** (2) 205-209 (2000)
6. M.V.Fedorov, L.V.Belousov, V.L.Voeikov, T.A.Zenchenko, K.I.Zenchenko, E.V.Pozharskii, A.A.Konradov, S.E.Shnoll "Synchronous Changes in Dark Current Fluctuations in Two Separate Photomultipliers in Relation to Earth Rotation" Astrophysics & Space Science №1 pp.105-112 (2003)
7. S. E. Shnoll, I.A.Rubinshtejn, K. I. Zenchenko, V.A.Shlekhtarev, A.V.Kaminsky, A.A.Konradov, N.V.Udaltsova.: Experiments with rotating collimators cutting out pencil of alpha-particles at radioactive decay of $^{239}$Pu evidence sharp anisotropy of space (2005). http://arxiv.org/abs/physics/0501004
8. Simon E. Shnoll " Changes in Fine Structure of Stochastic Distributions as a Consequence of Space-Time Fluctuations' ; Progress in Physics, V 2, April, 2006 Pp39-45
9. Rumer Yu.B. and Ryvkin M.Sh. Thermodynamics, Statistical Physics and Kinetics. Moscow, 1997, p. 367
10. Kaminsky A.V. The Hidden Space-Time in Physics. The Quantum Magic v. 2(1), p. 1101-1125. http://www.quantmagic.narod.ru/volumes/VOL212005/p1101.pdf
11. Kaminsky A.V. An Artifact or Regularity? Preprint in the library of MSU Seminar on temporology. http://www.chronos.msu.ru/RREPORTS/rlibrary.zip (2005)
12. H.Gould, J.Tobochnik "Computer Simulation Methods. Application to Physical Systems" Addison-Wesley Publishing Company, v.1 ch. 7, pp 81 и v.2, ch. 12,16.( 1988)
13. Pavel V. Kurakin, George G. Malinetskii, "Toy quantum mechanics using hidden variables". Discrete Dynamics in Nature and Society,:2 (2004), 357 – 361 (2004)
14. Gerard 't Hooft DETERMINISM BENEATH QUANTUM MECHANICS1. arXiv:quant-ph/0212095 v1 16 Dec (2002)